\documentclass[%
 preprint,
 superscriptaddress,
 amsmath,amssymb,
 aps,
 prapplied,
]{revtex4-2}

\usepackage{graphicx}
\usepackage{dcolumn}
\usepackage{bm}
\usepackage{siunitx}
\usepackage{braket}
\usepackage{geometry}
\usepackage{appendix}

\begin{document}

\title{Measuring the ferromagnetic resonance cone angle via static dipolar fields using diamond spins}

\author{B. A. McCullian}
  \affiliation{School of Applied and Engineering Physics, Cornell University, Ithaca, NY 14853, USA}
\author{M. Chilcote}
  \affiliation{School of Applied and Engineering Physics, Cornell University, Ithaca, NY 14853, USA}
\author{H. Yusuf}
  \affiliation{Department of Physics, The Ohio State University, Columbus, OH 43210, USA}
\author{E. Johnston-Halperin}
  \affiliation{Department of Physics, The Ohio State University, Columbus, OH 43210, USA}
\author{G. D. Fuchs}
 \email{gdf9@cornell.edu}
  \affiliation{School of Applied and Engineering Physics, Cornell University, Ithaca, NY 14853, USA}
  \affiliation{Kavli Institute at Cornell for Nanoscale Science, Ithaca, NY 14853, USA}
 
\date{\today}

\begin{abstract}
We demonstrate quantitative measurement of the ferromagnetic resonance (FMR) precession cone angle of a micro-scale sample of vanadium tetracyanoethylene (V[TCNE]$_{x\sim 2}$) using diamond spins. V[TCNE]$_{x\sim 2}$ is a low-damping, low-magnetization ferrimagnet with potential for scalable spintronics applications. Our study is motivated by the persistent need for quantitative metrology to accurately characterize magnetic dynamics and relaxation. Recently, diamond spins have emerged as sensitive probes of static and dynamic magnetic signals. Unlike analog sensors that require additional calibration, diamond spins respond to magnetic fields via a frequency shift that can be compared with frequency standards. We use a spin echo-based approach to measure the precession-induced change to the static stray dipolar field of a pair of V[TCNE]$_{x\sim 2}$ discs under FMR excitation. Using these stray dipolar field measurements and micromagnetic simulations, we extract the precession cone angle. Additionally, we quantitatively measure the microwave field amplitude using the same diamond spins, thus forming a quantitative link between drive and response. We find that our V[TCNE]$_{x\sim 2}$ sample can be driven to a cone angle of at least 6$^{\circ}$ with a microwave field amplitude of only 0.53 G. This work highlights the power of diamond spins for local, quantitative magnetic characterization.
\end{abstract}

\maketitle

\section{Introduction}

Quantitative magnetic metrology allows researchers to gain a fundamental understanding of magnetic materials and devices. In ferromagnetic resonance (FMR), a key parameter is the precession cone angle. This angle is determined by a balance between excitation and relaxation rates. Saturation of the cone angle~\cite{Olson_2007} can signal the onset of nonlinear effects such as Suhl instabilities~\cite{Suhl_1957}. Many spintronic devices, for example auto-oscillators~\cite{Kiselev_2003, Zahedinejad_2021}, rely on the precession of the magnetic moment. The cone angle contributes to the frequency of auto-oscillation~\cite{Slavin_2008}, the generated voltage in spin pumping~\cite{Costache_2008}, and lowered energetic thresholds for microwave-assisted switching~\cite{Nembach_2007}. 

Although the cone angle can be inferred from conventional techniques cavity FMR and broadband FMR, it is challenging to measure. Changes in magneto-resistance during precession can be used to quantify the cone angle~\cite{Costache_2006, Moriyama_2009, Kuhlmann_2012} but this requires fabrication of the magnetic material into a device. Optical techniques using the magneto-optical Kerr~\cite{Gerrits_2007} or Faraday~\cite{Bahlmann_1996} effects can also measure the cone angle, but these require optical access to the magnetic layer. Likewise, X-ray magnetic circular dichroism can probe the cone angle~\cite{Bauer_2015}, but such measurements require a beamline facility. Finally, all of these techniques require careful microwave calibration to quantify the microwave excitation amplitude at the sample position, which is notoriously imprecise due to transmission line standing waves.

Recently, sensitive measurements of magnetization dynamics have been demonstrated using nitrogen-vacancy (NV) center spins in diamond. These atomic-like sensors measure magnetic field changes as a frequency shift that can be compared with frequency standards, unlike analog detection techniques that require additional calibration. NV centers have a long-lived electronic spin state that can be optically initialized, manipulated by microwaves, and detected via fluorescence across a wide range of temperatures. NV centers can detect magnetic fields and noise spanning a frequency range from DC up to a few gigahertz. NV centers have been used to sense magnetic dynamics via both resonant coherent coupling~\cite{Andrich_2017, Kikuchi_2017, Bertelli_2020, Zhou_2021} and incoherent noise sensing~\cite{Du_2017, McCullian_2020, Koerner_2022, Solyom_2023, Fukami_2024} in a wide variety of magnetic systems. Such techniques also work for buried structures~\cite{Purser_2020}. Furthermore, magnetization dynamics can be sensed through changes in the static stray field of a sample when it undergoes precession~\cite{van_der_Sar_2015}. This technique relaxes the need for frequency matching between the magnetic dynamics and the NV center; the ferromagnet can be manipulated at any desired frequency, and the NV center senses changes of the static stray field. This response is directly related to the cone angle.

Here, we use an ensemble of NV centers in a diamond substrate to measure the uniform mode FMR of a micro-scale sample of vanadium tetracyanoethylene (V[TCNE]$_{x \sim 2}$). V[TCNE]$_{x \sim 2}$ is a ferrimagnet with low magnetization and damping~\cite{Yu_2014, Liu_2018, Franson_2019, Liu_2020, Cheung_2021, Yusuf_2021, Trout_2022} that can be deposited on a wide variety of substrates~\cite{Pokhodnya_2000, de_Caro_2000, Pokhodnya_2000, Harberts_2015, Zhu_2016}. In addition to its low damping ~\cite{Zhu_2016}, V[TCNE]$_{x \sim 2}$ is of interest for spintronic applications because of the presence of spin pumping~\cite{Liu_2018} and its ability to be patterned without sacrificing the sharp FMR linewidth~\cite{Franson_2019}, which has led to demonstrations of strong on-chip coupling between V[TCNE]$_{x \sim 2}$ and a superconducting circuit~\cite{Xu_2024} relevant to future work in quantum magnonics. In this study we detect the stray static magnetic field from the resonantly excited V[TCNE]$_{x \sim 2}$ using a modified spin-echo scheme. We first use the NV centers to spectroscopically characterize the uniform mode FMR condition. We then quantify the stray static field and relate changes of this field to the cone angle. We record the cone angle versus microwave field amplitude, finding that the V[TCNE]$_{x \sim 2}$ in our sample is driven to a cone angle of at least 6$^{\circ}$ at a microwave field amplitude of only 0.53 G.

\section{V[TCNE]$_{x \sim 2}$ disk pair on diamond}

\begin{figure}
\includegraphics{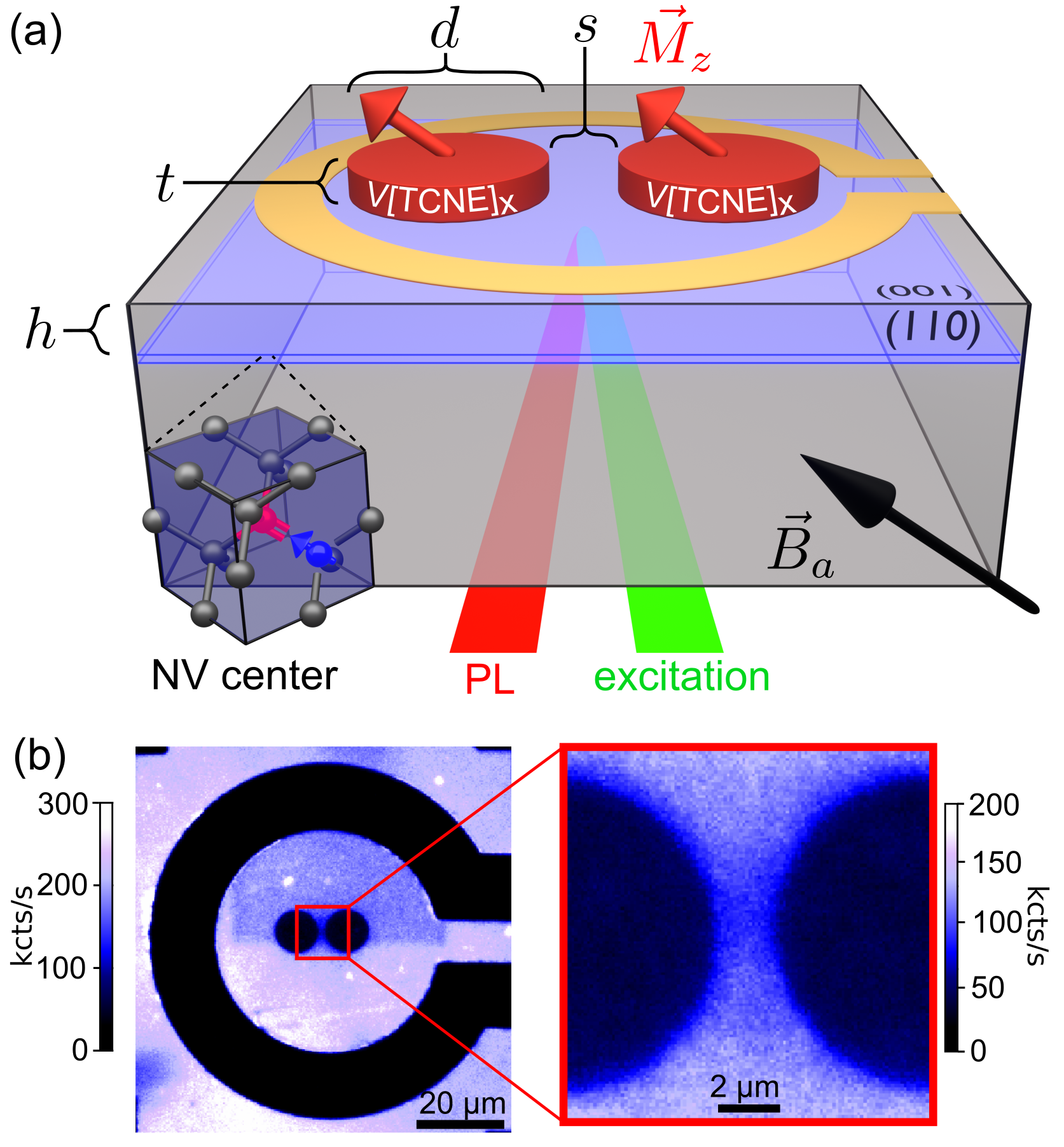}
\caption{\label{fig:Fig 1} V[TCNE]$_{x \sim 2}$ disk pair on diamond. (a) A V[TCNE]$_{x \sim 2}$ disk pair (red) with thickness $t$ = 700 nm, diameter $d$ = 9.5 $\unit{\um}$, and separation $s$ = 1.5 $\unit{\um}$ deposited inside a microwave antenna (gold loop) on the surface of a 50 $\unit{\um}$ thick diamond membrane (gray slab). From below, we optically excite and collect photoluminescence from an NV center ensemble implanted $h$ = 50 nm the below the diamond surface. An applied magnetic field $\vec{B}_a$ oriented near the diamond $\langle 1 1 1 \rangle$ crystal axis saturates the disks' magnetization $\vec{M}_z$. (b) Scanning PL images of the NV center ensemble show the disk pair and microwave antenna.}
\end{figure}

Our sample consists of a pair of V[TCNE]$_{x \sim 2}$ disks grown on diamond, as illustrated in Figure \ref{fig:Fig 1}(a). A pair is chosen because this geometry provides a more uniform stray dipolar field than a single disk. The substrate is a 50-micron-thick diamond membrane with an implanted NV center ensemble set $h = $ 50 nm beneath the surface. A microwave loop antenna (Ti [10 nm]/Pt [300 nm]) is fabricated on the diamond surface. This antenna is used both for driving electron spin resonance of the NV centers and for exciting ferromagnetic resonance in the V[TCNE]$_{x \sim 2}$ disks. Using low-temperature chemical vapor deposition through an electron-beam defined mask\cite{Yu_2014,Harberts_2015,Franson_2019}, we fabricate a disk pair inside the center of the microwave antenna. The disks have thickness $t$ = 700 nm, diameter $d$ = 9.5 $\unit{\um}$, and disk edge separation $s$ = 1.5 $\unit{\um}$. The sample was encapsulated with UV-cured epoxy\cite{Froning_2015} and a glass coverslip to prevent sample degradation. See Appendix \ref{sec:Appendix A} for additional sample fabrication details.

We mount the sample in a cryostat that maintains a fixed temperature of 200 K and optically address the NV centers using our home-built confocal photoluminescence (PL) microscope. An applied magnetic field $\vec{B_a}$ oriented near one of the diamond $\langle 111 \rangle$ axes selects one of the four families of NV center orientations in our diamond chip. Additionally, this applied field saturates the V[TCNE]$_{x \sim 2}$ magnetization $M_z$ along the applied field direction. Laser excitation and photoluminescence collection of the NV center ensemble are performed through the back surface of the diamond membrane. Scanning confocal microscopy images of the disk pair shown in Figure \ref{fig:Fig 1}(b) and cavity FMR measurements of a V[TCNE]$_{x \sim 2}$ growth witness film (see Appendix \ref{sec:Appendix A}) confirm the successful sample fabrication.

\section{Sensing FMR via stray dipolar fields}

\begin{figure}[t]
\includegraphics[scale=1]{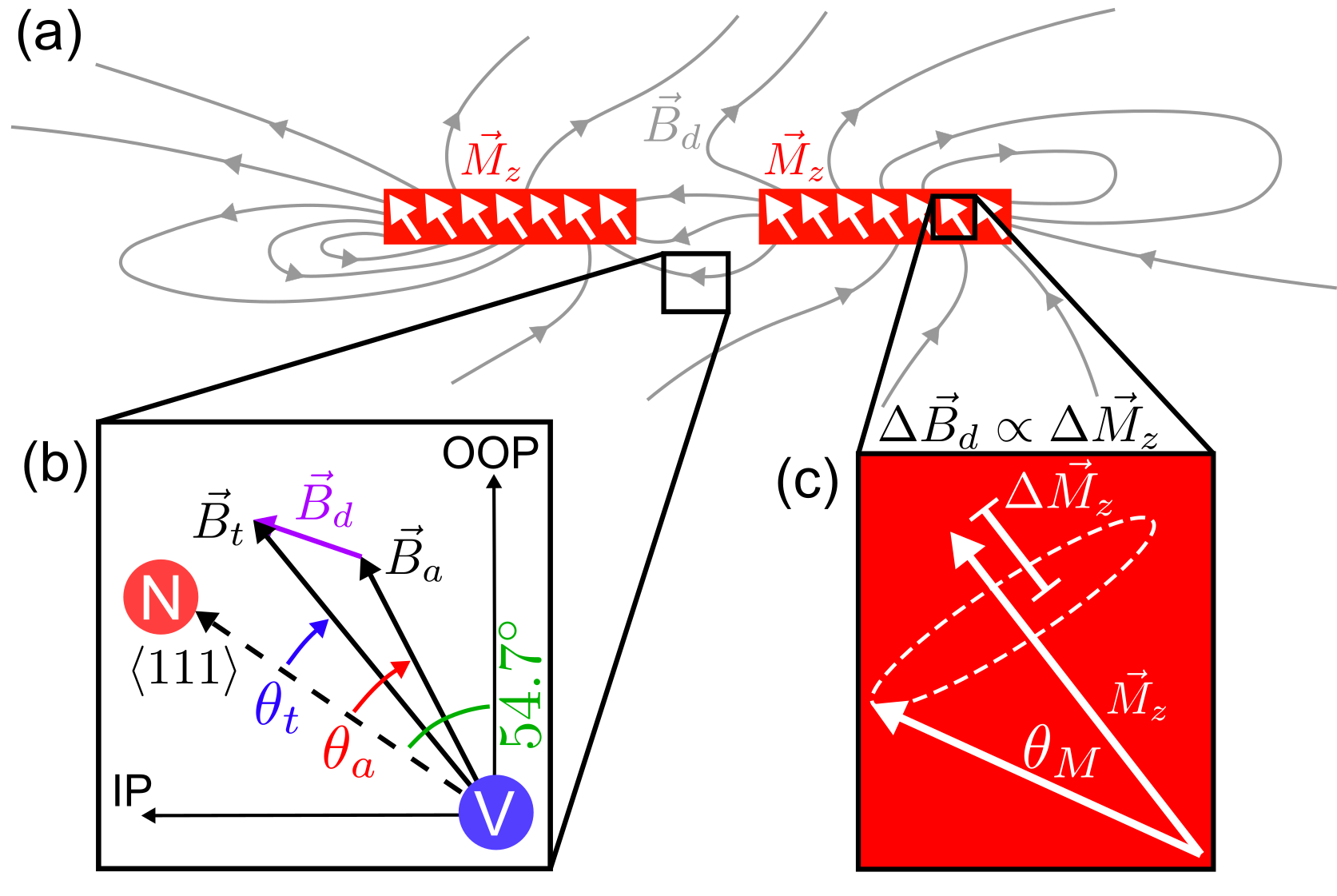}
\caption{\label{fig:Fig 2} Sensing the cone angle via static dipolar fields. (a) The disk pair magnetization $\vec{M_z}$ produces a static dipolar field $\vec{B_d}$ profile (gray arrows). (b) The total field $\vec{B_t}$ experienced by the NV center ensemble beneath the center of the disk pair is the vector sum of the applied field $\vec{B_a}$ and the dipolar field $\vec{B_d}$. We name the angle between $\vec{B_t}$ ($\vec{B_a}$) and the NV center symmetry axis as $\theta_t$ ($\theta_a$). (c) When FMR is excited in the disks, the magnetization precesses at a cone angle $\theta_M$, resulting in a proportional change of the static $\vec{B_d}$ experienced by the NV center ensemble.}
\end{figure}

We sense changes in the stray dipolar field of the magnetic disk pair during FMR excitation using the nearby NV center ensemble. The experimental geometry and the sensing mechanism are shown in Figure \ref{fig:Fig 2}. The static magnetization of the disks produces a static stray dipolar field ($\vec{B_d}$). The NV center ensemble beneath the disk pair experiences a total magnetic field ($\vec{B_t}$), which is the vector sum of $\vec{B_a}$ and $\vec{B_d}$. When the microwave field excites the disk pair into ferromagnetic resonance, changes in $\vec{M_z}$ cause a proportional change of $\vec{B_d}$. The precession cone angle of the magnetization is given by $\theta_M = cos^{-1}(1-\Delta M_z/M_z)$, where $\Delta M_z$ is the change in the magnetization along the equilibrium direction. This same relation holds for the dipolar field projected along the NV center axis ($B_{d,NV}$). By measuring the change of the dipolar field along the NV-center axis ($\Delta B_{d,NV}$) we can solve:
\begin{equation}
\label{eqn: Eqn 1}
\theta_M = cos^{-1}(1-\delta B_{d,NV}/B_{d,NV})
\end{equation}
to find the cone angle.

\section{FMR Echo Detection}

\begin{figure}[t]
\includegraphics[scale=1]{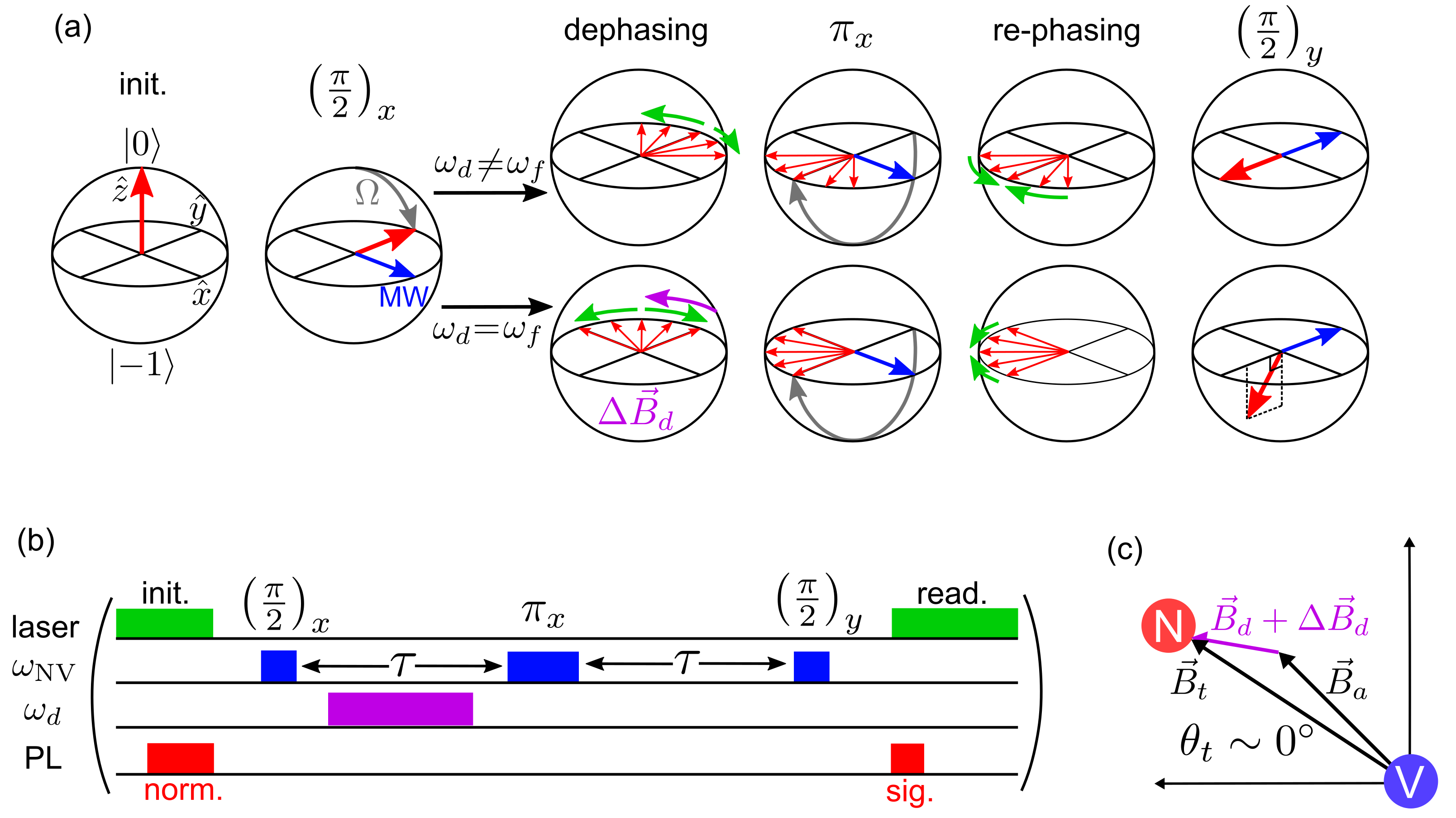}
\caption{\label{fig:Fig 3} FMR-echo sensing scheme (a) A Bloch sphere representation of the FMR-echo protocol. The NV centers are optically initialized into $\ket{0}$ and then coherently manipulated with microwaves at the NV center frequency $\omega_{\textrm{NV}}$. We use a $\big( \frac{\pi}{2} \big)_x - \tau - \big( \pi \big)_x - \tau - \big( \frac{\pi}{2} \big)_y$ echo sequence with the final pulse phase set to rotate about the $\hat{y}$ axis. An additional microwave drive at $\omega_d$ is applied during the dephasing window which is far detuned from $\omega_{\textrm{NV}}$. When $\omega_d \neq \omega_f$ (top row), no change in $\vec{B_d}$ occurs. This results in a final NV-center spin projection along the $\hat{-y}$ axis. However, when $\omega_d = \omega_f$ (bottom row), changes in $\vec{B_d}$ (purple arrow) cause the NV centers to acquire additional phase. This results in a final spin state projection with some component along the $\hat{z}$ axis. (b) The laser, microwave, and photon counting pulses used for FMR-echo detection. (c) At low microwave amplitudes, changes to $\vec{B_d}$ do not significantly affect $\theta_t$. The phase accumulated during the FMR drive is proportional to the change in the component of $\vec{B_d}$ along the NV-center symmetry axis.}
\end{figure}

Our measurements probe changes in $\vec{B_d}$ using a modified spin echo sensing scheme. We refer to this scheme as an FMR-echo. We choose an echo sequence rather than a Ramsey sequence because the echo removes the effects of magnetic field inhomogeneity arising from dipole field gradients across the NV-center ensemble. Figure \ref{fig:Fig 3}(a) shows the NV-center response to the FMR-echo sequence. The corresponding experimental pulses are shown in Figure \ref{fig:Fig 3}(b). We work in the NV-center spin state basis $\{\ket{m_s=0},\ket{m_s=-1} \}$, which we refer to as $\ket{0}$ and $\ket{-1}$. These states have an energy splitting of $\hbar \omega_{\textrm{NV}}$. A 532 nm, 30 $\unit{\uW}$ laser applied for 30 $\unit{\us}$ initializes the NV-center spins to $\ket{0}$. The NV centers are coherently driven through a spin echo sequence where the rotations of each pulse are given by: $\big( \frac{\pi}{2} \big)_x - \tau - \big( \pi \big)_x - \tau - \big( \frac{\pi}{2} \big)_y$ with delay times $\tau$, and where the subscript describes the rotation axis in the Bloch sphere. Crucially, the final microwave rotation carries a phase such that spin rotation is about the $y$-axis, making the final NV-center spin projection linearly sensitive to small changes in $\Delta B_d$. An additional microwave pulse at a drive frequency $\omega_d$, far detuned from $\omega_{\textrm{NV}}$, is applied during the dephasing window to excite FMR in the disk pair.

The FMR-echo signal corresponds to the final spin state projection of the NV centers onto the $z$-axis. When $\omega_d$ is not resonant with the FMR condition of the disks ($\omega_f$), the final NV-center spin state has a zero projection along the $z$-axis. However, when $\omega_d = \omega_f$ the NV centers accumulate additional phase during the FMR drive. This additional phase accrual results in a final spin projection with a non-zero component along the $z$-axis that is proportional to the change in $\vec{B_d}$ along the NV-center symmetry axis.

\section{FMR Echo Spectroscopy}

\begin{figure}[t]
\includegraphics[scale=1]{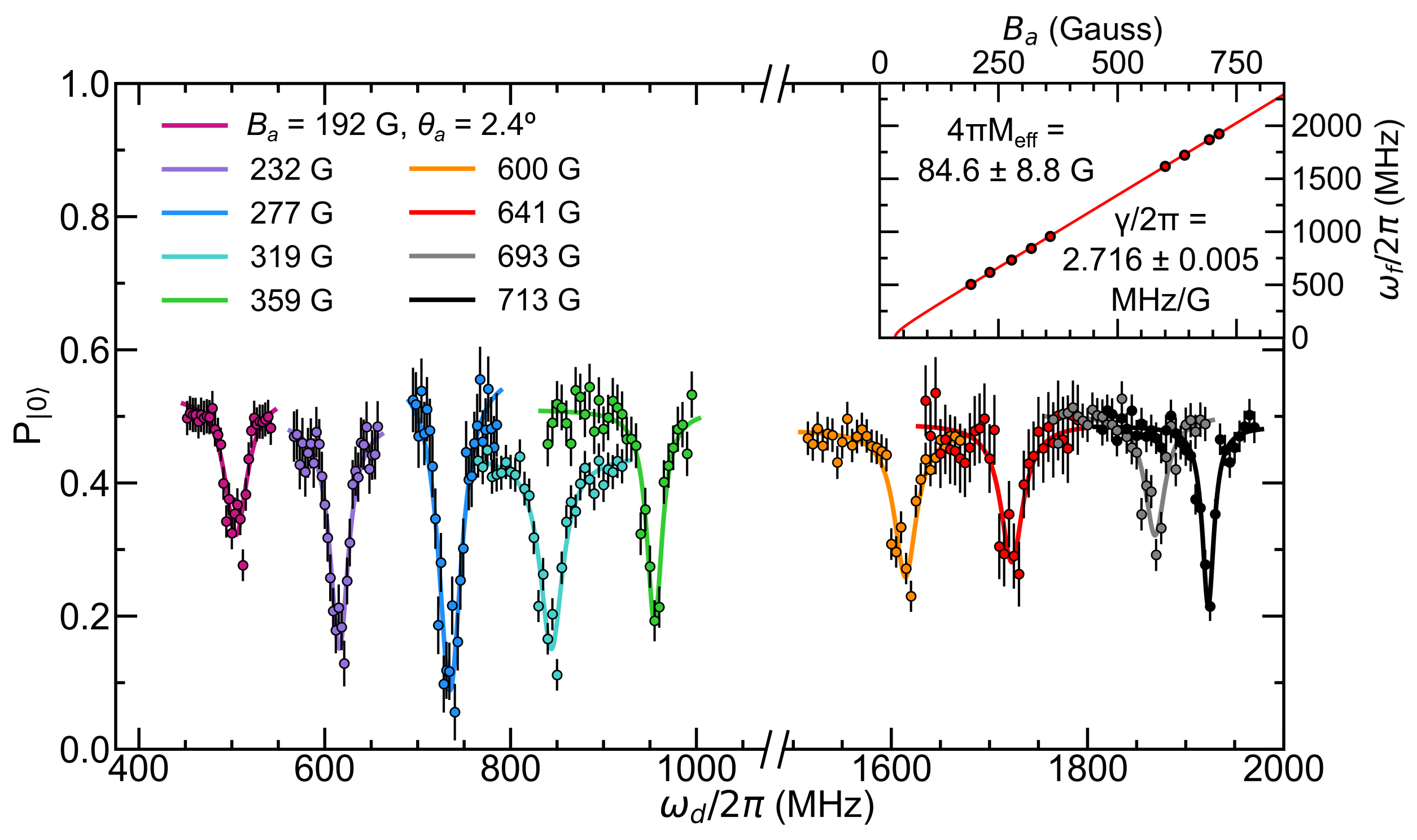}
\caption{\label{fig:Fig 4} FMR-echo spectroscopy of V[TCNE]$_{x \sim 2}$ disk pair. FMR-echo signal versus $\omega_d$ for several discrete values of $\vert \vec{B_a} \vert$. Vertical error bars assume a photon shot noise limited detection. $\theta_a$ is set to 2.4$^\circ$ such that $\theta_t \approx 0^\circ$ (see Appendix~\ref{sec:Appendix B}). The resonance frequency for each applied field is extracted by fitting a Lorentzian response (solid curves). (inset) The FMR condition in field and frequency. Fitting this response to Equation \ref{eqn: Eqn 2} yields an effective magnetization of $4 \pi M_{\textrm{eff}} = 84.6 \pm 8.8$ G and a gyromagnetic ratio of $\gamma / 2 \pi = 2.716 \pm 0.005$ MHz/G, consistent with previous results~\cite{Plachy_2004,Zhu_2016,Franson_2019,Yusuf_2021, Bola_2021}}
\end{figure}

We measure the FMR dispersion of our disk pair using the FMR-echo sequence. The applied field is oriented at an angle such that the total magnetic field $\vec{B_t}$ is closely aligned to the NV center axis, as shown in Figure \ref{fig:Fig 3}(c). The applied magnetic field angle was set so that the total magnetic field aligned with the NV-center symmetry axis near the excited-state avoided level crossing at about 500 G~\cite{Epstein_2005}. Using mumax3 micromagnetic simulations~\cite{Vansteenkiste_2014}, we determined the components of the applied and dipolar magnetic fields (See Appendix \ref{sec:Appendix B} for details). The applied field angle is $\theta_a = 2.4^\circ$. We note that $\vec{B_d}$ and $\vec{B_a}$ are in general not collinear. We verified this with additional continuous-wave measurements of the disk pair FMR using the NV center ensemble (see Appendix \ref{sec:Appendix B2}). For the FMR-echo sequence, $\tau$ is 700 ns and the FMR drive pulse duration is 680 ns. FMR-echo signal is averaged for 250 seconds at each frequency, corresponding to $7.6 \times 10^6$ repetitions of the FMR-echo sequence at each point. We recorded the final spin state projection of the FMR-echo sequence versus the drive frequency, $\omega_d$. The results taken at several values of the applied field magnitude are shown in Figure \ref{fig:Fig 4}. 

The FMR condition is given by~\cite{Franson_2019}:
\begin{equation}
\label{eqn: Eqn 2}
\omega_f = \gamma \sqrt{\Bigl(B_a - 4 \pi M_{\textrm{eff}}\cos^2(\theta)\Bigr) \Bigl( B_a - 4 \pi M_{\textrm{eff}} \cos (2 \theta ) \Bigr)}
\end{equation}
where $\omega_{f}$ is the FMR resonance frequency, $\gamma$ is gyromagnetic ratio for V[TCNE]$_{x \sim 2}$, $4 \pi M_{\textrm{eff}}$ is the V[TCNE]$_{x \sim 2}$ effective magnetization, and $\theta$ is the applied magnetic field angle. For our measurements, the applied field is tilted further towards the out-of-plane direction than the NV-center bond axis direction, such that $\theta$ = 52.3$^{\circ}$. For each value of the applied magnetic field in Figure \ref{fig:Fig 4} we fit a Lorentzian response and extract the resonance frequency. We then fit the field-frequency dispersion to Equation \ref{eqn: Eqn 2}, as shown in the inset of Figure \ref{fig:Fig 4}. We find that $4 \pi M_{\textrm{eff}}$ = 84.6 $\pm$ 8.8 G and $\gamma/2\pi$ = 2.716 $\pm$ 0.005 MHz/G, both of which are consistent with values reported in the literature~\cite{Plachy_2004,Zhu_2016,Franson_2019,Yusuf_2021, Bola_2021}. We note that our measurement of the FMR dispersion did not require frequency matching between the NV-center spin resonance and the V[TCNE]$_{x \sim 2}$ disks.

\section{Cone Angle Measurements}

\begin{figure}[t]
\includegraphics[scale=1]{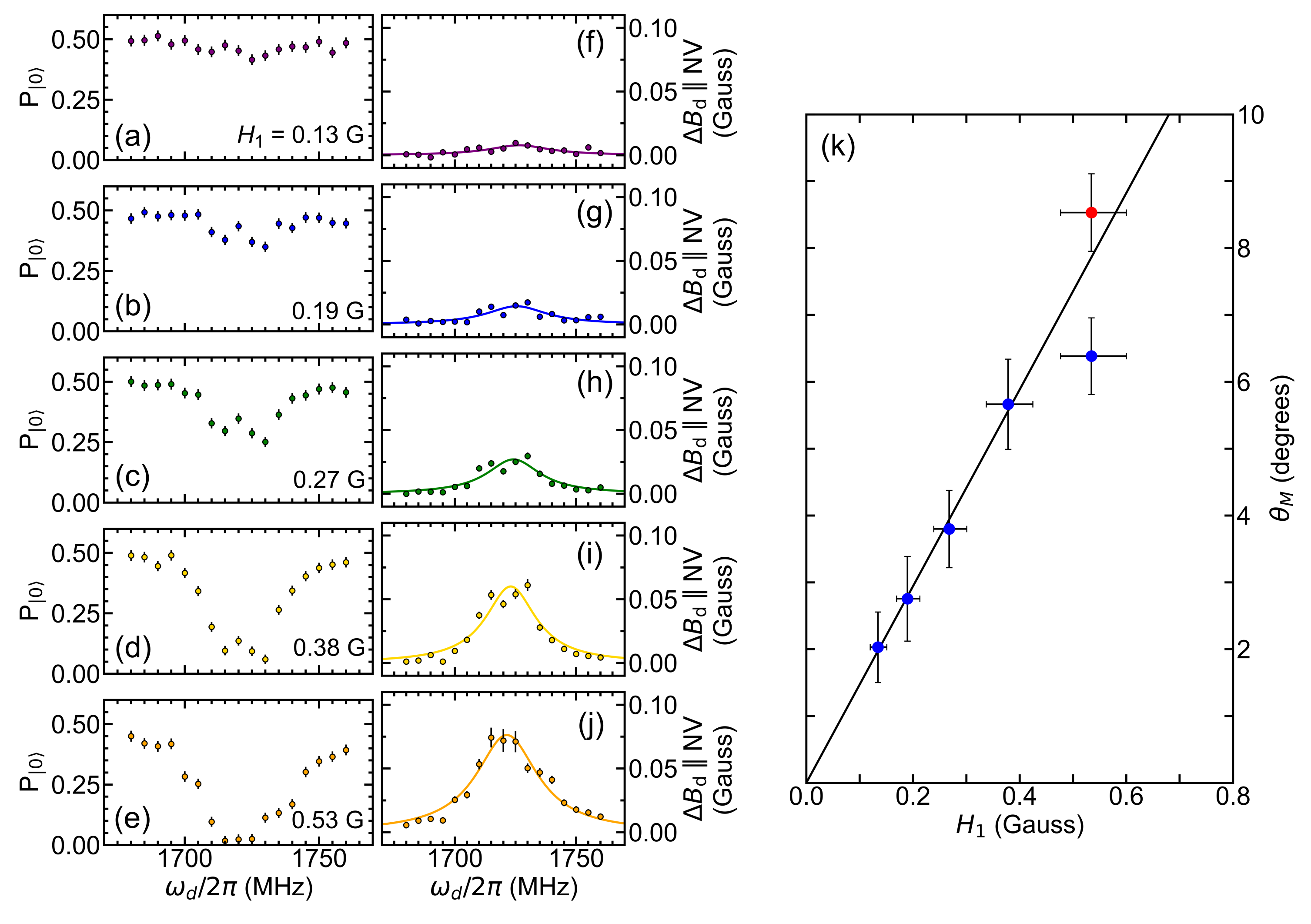}
\caption{\label{fig:Fig 5} Cone angle sensing using the FMR-echo protocol. (a-e) FMR-echo signal versus $\omega_d$ at several $H_1$ amplitudes. The vertical error bars assume photon shot noise limited detection. (f-j) Calculated $\Delta B_{d,NV}$ versus $\omega_d$. Lorentzian fits (solid curves) are used to extract the maximum sensed field for each value of $H_1$. (k) The maximum calculated cone angle $\theta_M$ for each $H_1$ (blue circles). Vertical error bars are determined by both the fit amplitude uncertainty in panels (f-j) and a $\pm$10\% uncertainty in the saturation magnetization of the discs. Horizontal error bars correspond to a $\pm$ 1 dB uncertainty in our applied microwave power. For $H_1 = $0.53 G, we accrue sufficient phase during the FMR-echo that there is an ambiguity in assigning a phase to the data in panel (e). The red data point in panel (k) assumes that the lowest three data points in panel (e) have accrued more than 90$^\circ$ of phase. The solid line is a linear fit to the first four blue data points.}
\end{figure}

Using the FMR-echo scheme, we measure the precession cone angle. We record sweeps of the FMR-echo signal versus drive frequency for several values of the microwave amplitude as shown in Figure \ref{fig:Fig 5}(a-e). The microwave amplitudes were determined from separate measurements in which we directly drove Rabi nutations of the NV center ensemble away from the disks, at magnetic field values corresponding to the NV center spin resonance and not the FMR condition.  This gives us the microwave amplitude at a particular frequency and power without assumption about microwave transmission properties. We convert the FMR-echo signals into a changes in the dipolar field along the NV-center axis in Figure \ref{fig:Fig 5}(f-j). We first convert the spin-state projection signal from the FMR-echo into a phase accrued on the NV centers during the FMR driving pulse ($\sin(\phi) = 2(P_{\ket{0}}-0.5)$) and then relate that phase to a change in the dipolar field along the NV-center axis ($\phi = \gamma \tau \Delta B_{d,NV}$)~\cite{Barry_2020}. We fit a Lorentzian response to the extracted $\Delta B_{d,NV}$ versus drive microwave frequency for each microwave amplitude.

The cone angle versus the microwave amplitude is plotted in Figure \ref{fig:Fig 5}. The cone angle for each microwave amplitude is determined by converting the peak values of $\Delta B_{d,NV}$ in Figure \ref{fig:Fig 5}(f-j) to $\theta_M$ using Equation \ref{eqn: Eqn 1}. We note that assigning a phase to the peak data in Figure \ref{fig:Fig 5}(e) is ambiguous because for NV-center phase accumulations near 90$^{\circ}$ the final pulse in our sequence will rotate the spins toward $\ket{-1}$. Thus, we also report the red data point in Figure \ref{fig:Fig 5}(k) that assumes a phase of more than 90$^{\circ}$ has accumulated. A linear response fit to the four lowest microwave amplitudes indicates the expected response of the V[TCNE]$_{x \sim 2}$ cone angle in the low-microwave-power regime~\cite{Fan_2010}. Because of the ambiguity in our measurement we are unable to say if this linear trend continues with the red data point, or if the cone angle begins to saturate at onset of nonlinear effects~\cite{Suhl_1957}.  The V[TCNE]$_{x \sim 2}$ cone angle can be driven to at least 6$^{\circ}$ at a microwave field amplitude of 0.53 G. We conclude from this that the low damping in V[TCNE]$_{x \sim 2}$ allows for the magnetization to be driven to large cone angles at modest microwave fields as compared to materials such as permalloy~\cite{Costache_2008}.

\section{Discussion}

We have demonstrated that magnetization dynamics of micro-scale magnets can be probed by measuring changes in the stray static magnetic field of a precessing magnet with nearby NV centers. Using FMR-echo, we measured the V[TCNE]$_{x \sim 2}$ FMR condition, finding that the deposition and patterning on diamond left the magnetization and gyromagnetic ratio as expected. We recorded the V[TCNE]$_{x \sim 2}$ cone angle versus microwave amplitude and found an expected linear dependence of the cone angle at small microwave field amplitudes. That V[TCNE]$_{x \sim 2}$ can be driven to a cone angle of at least 6$^{\circ}$ with less than 1 G of microwave field highlights an opportunity: a simple probe of the cone angle can aid in the search for useful materials for low-power spintronics applications.

The FMR-echo technique is non-invasive, requires no electrical contact or direct optical access to the magnetic layer, and can be operated using table-top optical and microwave equipment. The long signal integration times used in our measurement were the result of working at low laser power to avoid damaging the V[TCNE]$_{x \sim 2}$~\cite{Cheung_2021}. For solid-state magnets the measurements can be significantly faster. We expect that, given the modest time delays used in our FMR-echo sequence, scanning probe instruments with diamond tips could perform similar measurements, with the added benefits of positioning flexibility and a measurement at a well-defined point in space. Measuring a material with higher saturation magnetization than V[TCNE]$_{x \sim 2}$, for instance permalloy, will introduce significant magnetic field inhomogeneity. This can be overcome by using single NV-center measurements at the cost of longer integration times, or by working further from the discs since the magnetic field gradient decreases faster than the field. Previously, a similar echo-based technique was used to probe the uniform mode dynamics of permalloy using a single NV center\cite{van_der_Sar_2015}. Thus, the FMR-echo technique is a good probe of magnetization dynamics across a wide range of magnet properties, including saturation magnetization, damping, and electrical conductivity. Relaxometry approaches to measuring magnetization dynamics require exciting spin noise at the NV-center frequency~\cite{Wolfe_2014, van_der_Sar_2015, Du_2017, McCullian_2020}, which can limit experiments to low frequencies. Because the FMR-echo does not require such matching, it can be used to measure high frequency dynamics, for example in antiferromagnetic resonance in materials with a canted moment~\cite{Boventer_2021}.

In future work ambiguities in mapping a final spin state population to a phase can be resolved by using two measurements, one each around the $x$- and $y$-axes for the final readout pulse. This will allow for extension of our measurements to higher microwave drive powers in V[TCNE]$_{x \sim 2}$ and would additionally be useful for magnets with higher saturation magnetization, where the changes in the stray static field can be much larger. Measuring at higher microwave amplitude will allow characterization of the onset of nonlinear effects. Because NV centers can measure both DC and AC magnetic field, they can be used to measure the applied field, the static dipolar field, and the microwave drive field. Although in this work we relied on micromagnetic simulations of the disk pair stray field to get the static dipolar field, future studies can directly probe these stray fields. Finally, a verification of FMR-echo measurements of the cone angle against magnetoresistance measurements will be useful for benchmarking FMR-echo as a tool for high-precision cone angle measurements.

\section{Acknowledgments}

The design and fabrication of our device, all the measurements, and all data analysis were supported by the Department of Energy Office of Science, Basic Energy Sciences Quantum Information Sciences program (DE-SC0019250). The diamond substrate and microwave antenna fabrication made use of facilities at the Cornell NanoScale Facility, an NNCI member supported by the NSF (NNCI-2025233) and the Cornell Center for Materials Research Shared Facilities which were supported through the NSF MRSEC program (DMR-1719875). For the V[TCNE]$_x$ disc fabrication, the authors acknowledge partial support from the NanoSystems Laboratory User Facility supported by the Center for Emergent Materials, an NSF MRSEC (DMR-2011876).

\clearpage

\begin{appendix}

\counterwithin{figure}{section}
\setcounter{figure}{0}

\section{Sample Fabrication}\label{sec:Appendix A}

\subsection{Diamond substrate fabrication}

Our starting substrate is a $\approx$50-$\unit{\um}$-thick type IIa diamond membrane with \textless 1 part-per-billion nitrogen impurities.  The top 5~\unit{\um} was removed by alternating O\textsubscript{2} and Ar/Cl plasma etching \cite{Chen_Nanoletters_2019} to release surface stress due to polishing. The diamond was implanted with nitrogen ions and subsequently annealed at 1100$^\circ$ C for 2 hours to form an ensemble of NV centers at a depth of 50 nm below the surface. The sample was then cleaned in a heated mixture of nitric, sulfuric, and perchloric acid. Using optical lithography and lift-off techniques we fabricated microwave loop antennae (Ti (10 nm)/Pt (300 nm)) on the diamond surface for microwave spin control. 

\subsection{V[TCNE]$_{x \sim 2}$ disk pair fabrication}

Given the relatively low temperature and high pressure deposition conditions, it is possible to integrate patterned V[TCNE]$_{x \sim 2}$ with preexisting, on-chip device structures such as microwave waveguides and alignment markers using electron-beam lithography. This approach relies on two innovations unique to V[TCNE]$_{x \sim 2}$ patterning~\cite{Franson_2019}: (i) the identification of a solvent, dichloromethane, that removes electron-beam resist, but does not negatively impact V[TCNE]$_{x \sim 2}$ magnetic properties, and (ii) the inclusion of a thin Al$_2$O$_3$ layer ($2-3$ nm) above the resist that prevents residual solvent from contaminating the V[TCNE]$_{x \sim 2}$ layer. In addition to robust on-chip integration, this approach preserves the coherence properties of magnonic excitations within the patterned structures~\cite{Franson_2019}.

\begin{figure}[t]
\includegraphics[scale=1]{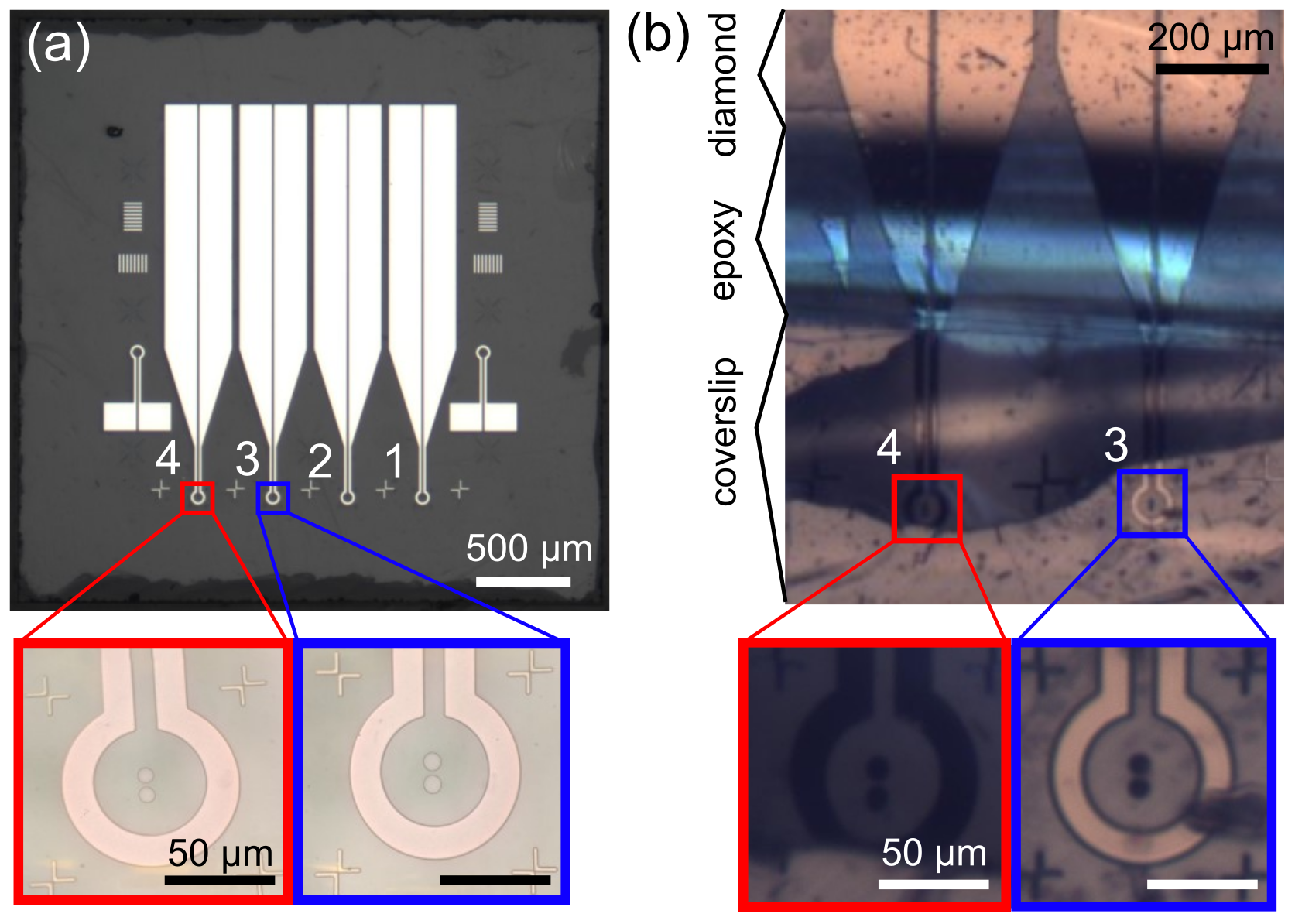}
\caption{\label{fig:Fig S1} V[TCNE]$_x$ disk pair on diamond fabrication. (a) Optical micrographs of the diamond membrane with deposited microwave antennae and disk pair resist patterns before V[TCNE]$_{x \sim 2}$ deposition. We deposited four microwave loop antennae (Ti (10 nm)/Pt (300 nm)) on the surface of the diamond membrane following nitrogen implantation, annealing, and acid cleaning. Using electron-beam lithography, we created disk pair patterns inside the center of each loop. (b) Optical micrographs of the completed and encapsulated sample. We deposited $600$-$700$ nm thick V[TCNE]$_{x \sim 2}$ disk pairs via the standard low-temperature CVD process~\cite{Yu_2014,Harberts_2015,Chilcote_2019,Yusuf_2021}, followed by liftoff in dichloromethane~\cite{Franson_2019}. The pairs are encapsulated with a UV-cured epoxy~\cite{Froning_2015} and a piece of glass coverslip, such that the antenna electrodes are accessible for wire-bonding to our cryostat sample holder.}
\end{figure}

The $\sim 5 \times 5$ mm diamond membrane device (Figure \ref{fig:Fig S1}(a) is mounted on a $2 \times 2$ cm SiO$_x$ carrier wafer using PMMA as an adhesive. This allows for even deposition of the electron-beam resist during the spin-coating process. A 400 nm layer of MMA (8.5) MAA EL 11 (P(MMA-MAA)) is spun on at 2000 rpm for 45 seconds then soft baked at 180°C for 300 seconds. A 140 nm layer of 495PMMA A6 (PMMA) is then spun on at 2000 rpm for 45 seconds, then soft baked at 180°C for 60 seconds. A 10-nm-thick layer of aluminum is deposited via thermal deposition at 1$\times$10$^{-6}$ Torr. The electron-beam patterning of disk shapes on the PMMA/P(MMA-MAA) bilayer is performed on a FEI Helios Nanolab 600 Dual Beam Focused Ion Beam/Scanning Electron Microscope with the assistance of Nanometer Pattern Generation System (NPGS) software; a scanning magnification of 100 X with a beam area-dosage of 140.232 $\mu$C/cm$^2$ were used as optimal patterning parameters. Four sets of disk pairs are patterned in the center of the antenna loops; each disk has a diameter $<$ 10 $\mu$m (disks for antennae 1 and 2 have a diameter of about 7.37 $\mu$m, those for antenna 3 have a diameter of about 8.42 $\mu$m, while the disks for antenna 4 has a diameter of around 9.47 $\mu$m) with a disk separation of $<$ 2 $\mu$m (pairs 1 and 2 have a separation distance of about 1.63 $\mu$m, pair 3 of around 1.58 $\mu$m, while pair 4 has a separation distance of about 1.53 $\mu m$). Development of the written pattern is achieved with Kayaku MF-319 for 40 seconds, DI water for 20 seconds, Kayaku MIBK:IPA (1:3) for 60 seconds, IPA for 20 seconds, DI water for 20 seconds. That is followed by a 120 second hard bake at 100°C. A 3-nm-thick layer of aluminum is then deposited in the same system as the prior 10 nm layer to prevent outgassing of the PMMA/P(MMA-MAA) bilayer during V[TCNE]$_x$ deposition. The sample is oxidized and cleaned with a 10-minute UV-Ozone cleaning. This oxidizes the surface of the 3 nm aluminum layer and removes potential small sources of contamination from the growth surfaces.

The patterned substrates are transferred into the glovebox housing the CVD reactor, and a V[TCNE]$_{x \sim 2}$ film of $600$-$700$ nm is deposited using standard thin-film deposition protocols consistent with previous reports~\cite{Yu_2014,Harberts_2015,Chilcote_2019,Yusuf_2021}. Briefly, argon gas transfers the two precursors tetracyanoethylene (TCNE) and vanadium hexacarbonyl (V(CO)$_6$) into the reaction zone of a custom-built CVD reactor where V[TCNE]$_{x \sim 2}$ is deposited onto the diamond membrane. The system is temperature controlled to maintain the TCNE, V(CO)$_6$ and the reaction zones at 65°C, 10°C and 50°C respectively, which results in a typical growth rate of $\sim 200$-$300$ nm/hr. After growth, liftoff takes place in the glovebox to remove the bulk film so as to only leave the disk patterned V[TCNE]$_{x \sim 2}$. Liftoff is accomplished by immersion in dichloromethane for $\sim 10$ minutes accompanied by gentle agitation.

A completed membrane structure is strategically encapsulated with a UV-cured epoxy~\cite{Froning_2015} and a $\sim 1 \times 2.2$ mm piece of glass coverslip, such that the antenna electrodes are accessible for wire-bonding to measurement chip.

\begin{figure}[t]
\includegraphics[scale=1]{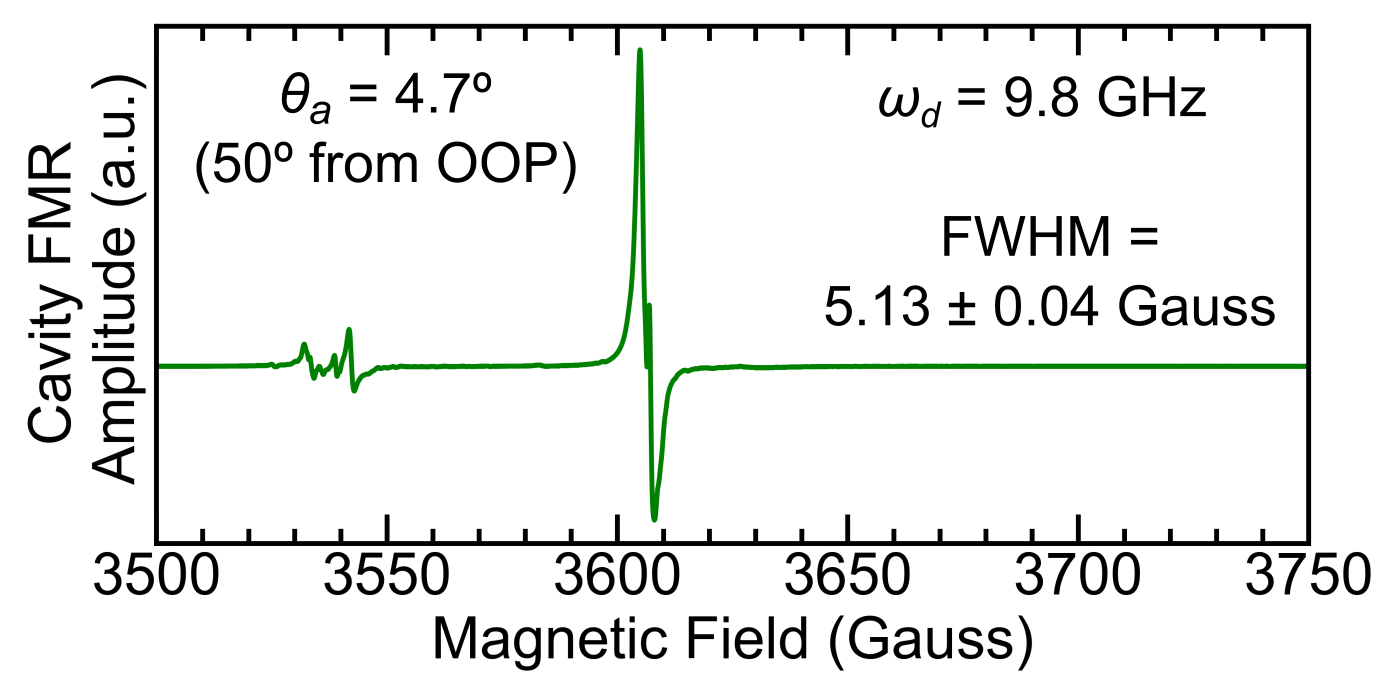}
\caption{\label{fig:Fig S2} A cavity FMR measurement of our V[TCNE]$_{x\sim 2}$ growth witness sample taken at $\omega_d = $9.8~GHz with the applied field 50$^\circ$ from the films out-of-plane direction, a similar angle to the NV-center measurements shows the quality of the film.}
\end{figure}

We measure the cavity FMR response at 9.8 GHz of an unpatterned witness film, deposited simultaneously to the disk pair sample, to verify the quality of the V[TCNE]$_{x \sim 2}$ growth. This measurement is taken with the static magnetic field oriented 50$^\circ$ from the out-of-plane direction, similar to the applied field angles probed in the experiments using the NV center ensemble for detection. The single-peaked response with sharp linewidth (Figure \ref{fig:Fig S2}) confirms the film quality. The full-width-at-half-maximum linewidth of 5.13 G is consistent reported linewidths in the literature \cite{Yu_2014, Liu_2020, Trout_2022}. By assuming that there are no inhomogeneous contributions to the linewidth, we can establish an upper bound on the magnetic damping for our V[TCNE]$_{x \sim 2}$ of $\alpha = 7.33 \pm 0.06 \times 10^{-4}$.

\section{Determining $\vec{B_d}$ and $\vec{B_a}$}\label{sec:Appendix B}

\subsection{Procedure for determining $\vec{B_a}$ and $\vec{B_d}$}\label{sec:Appendix B1}

Quantifying the changes in the disk pair's static dipolar field as measured by the NV center ensemble requires that we know the magnitude and direction of both $\vec{B_d}$ and $\vec{B_a}$. We determine these values for all measurements using the following procedure. First, we leverage the sensitivity of the NV centers' photoluminescence to the magnitude and direction of $\vec{B_t}$ under continuous laser illumination near the NV-center excited-state avoided level crossing (at approximately 512 G)~\cite{Epstein_2005} to set the distance and angle of our permanent magnet such that $\vec{B_t}$ is along the diamond $\langle 111 \rangle$ for $\vert \vec{B_t} \vert = 512$ G. Then, using micromagnetic simulations (see Appendix \ref{sec:Appendix B3} for details) of the $\vec{B_d}$ profile of the disk pair that incorporate the known magnetic properties of V[TCNE]$_{x \sim 2}$, we vary the magnitude and direction of the $\vec{B_a}$ in the simulation until we find the condition for which $\vec{B_t}$ is along the diamond $\langle 111 \rangle$ and $\vert \vec{B_t} \vert = 512$ G. This sets the applied magnetic field direction for all our NV experiments and simulations to 2.4$^\circ$ from the diamond $\langle 111 \rangle$ axis, toward the out-of-plane direction. For all other values of the applied magnetic field we vary the amplitude of $\vec{B_a}$, keeping the direction fixed, until we find an applied field magnitude such that the simulated $\vec{B_t}$ results in an NV center spin $\ket{m_s = 0} \leftrightarrow \ket{m_s = -1}$ transition frequency within 1.5 MHz of the measured NV center spin transition frequency. This corresponds to an error in the total magnetic field magnitude of about 0.5 G. The smallest total field used in the experiment is more than 200 G, thus for each measurement we know the full set of components of $\vec{B_a}$ and $\vec{B_d}$ to within 0.3\% error.

\subsection{Continuous-wave FMR angle sensing}\label{sec:Appendix B2}

\begin{figure}[t]
\includegraphics[scale=1]{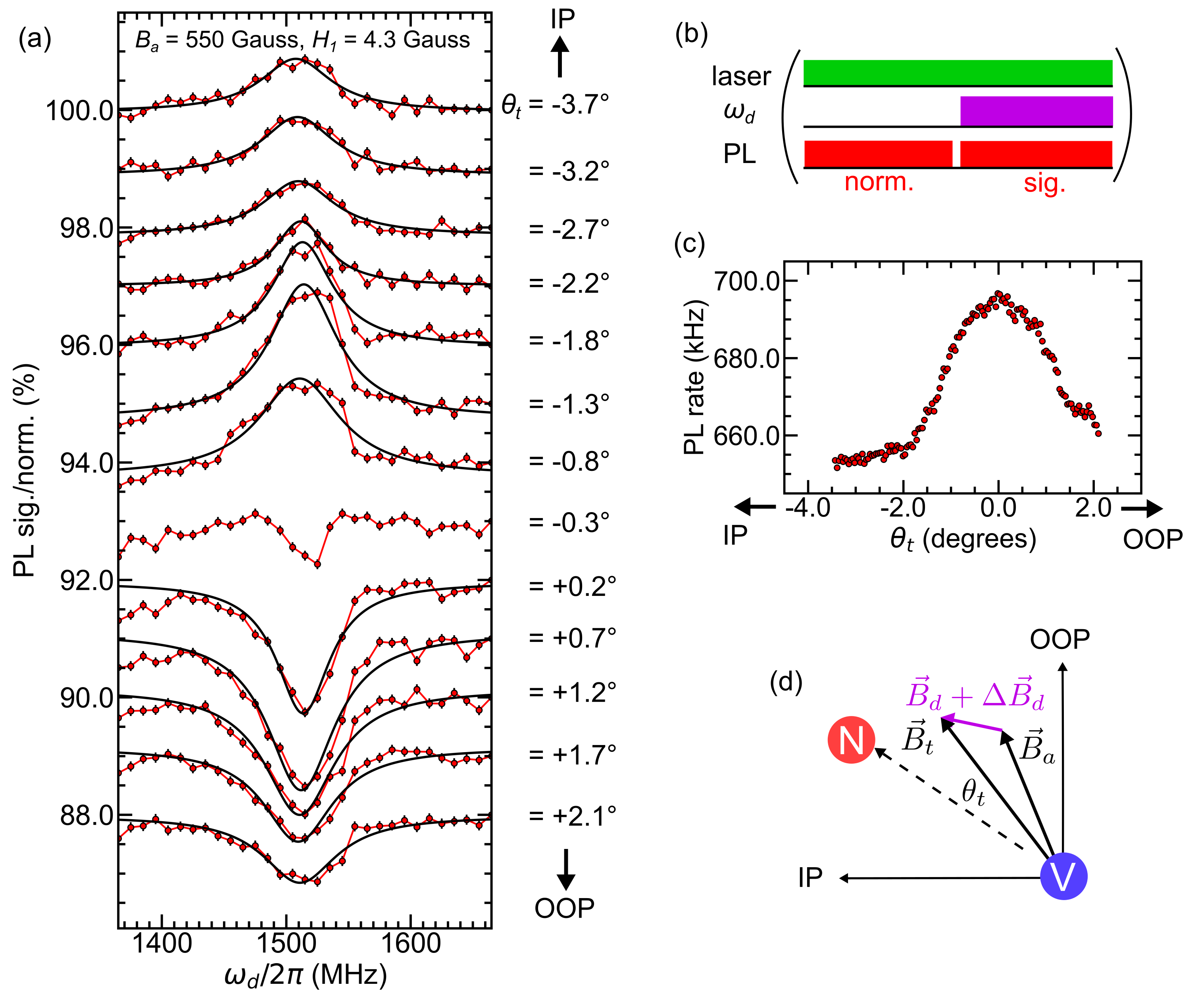}
\caption{\label{fig:Fig S3} Continuous wave sensing of resonantly-driven disk pair static stray field (a) Continuous-wave measurements of the NV center ensemble PL while sweeping $\omega_d$, which is far detuned from $\omega_{\textrm{NV}}$. Each curve is offset from its neighbor by 1\%. From micromagnetic simulations (see Appendix \ref{sec:Appendix B3}) we determine that $\vert \vec{B_a} \vert = 550$ G. We determine that $H_1 = 4.3$ G from NV center Rabi nutation measurements. The resonant response at $\omega_d = 1520$ MHz corresponds to the disk pair FMR condition. Measurement of the CW PL response at several values of $\theta_t$ shows inversion of the sign of the PL contrast. (b) The laser, microwave, and photon counting pulses used in the continuous-wave sequence. (c) The raw photoluminescence count rate of the NV center ensemble under continuous illumination versus $\theta_t$ shows PL quenching due to spin-state mixing induced by the off-axis magnetic field\cite{Epstein_2005}. (c) When the disk pair is driven to large values of the cone angle at high microwave amplitude, the resulting changes in $\vec{B_d}$ result in changes of $\theta_t$ by as much as a few degrees which results in the PL changes seen in panel (a).}
\end{figure}

We experimentally verify that $\vec{B_a}$ and $\vec{B_d}$ are not collinear using a simple, continuous-wave FMR detection scheme using the NV center ensemble. The results are shown in Figure \ref{fig:Fig S3}(a). We perform sweeps of the microwave drive frequency using a conventional CW ODMR scheme (\ref{fig:Fig S3}(b)) at an applied field of 550 G and with high microwave intensity (4.3 G, as determined by NV center Rabi nutation experiments). The resonant PL response at $\omega_d = 1520$ MHz is far detuned from any NV center resonances, ensuring no direct microwave drive of the NV center spin state. Instead, the ODMR contrast occurring at 1520 MHz and 550 G is consistent with the disk pair FMR condition. The observed contrast changes sign as we vary the angle of $\vec{B_a}$. We note that direct microwave drive of NV center spins under continuous laser illumination cannot itself account for a positive spin contrast.

We account for the sign change in the ODMR contrast on the FMR condition by considering the sensitivity of the NV center photoluminescence near the excited state avoided level crossing~\cite{Epstein_2005}. Under continuous illumination and with no microwave drive, the PL versus the applied field angle is measured, as shown in Figure \ref{fig:Fig S3}(c). We attribute the observed changes in PL found in \ref{fig:Fig S3}(a) to this effect. As shown in \ref{fig:Fig S3}(d), when the dipolar field of the disk pair is sufficiently reduced during FMR drive at high microwave intensity, the angle of $\vec{B_t}$ can change. At an applied field of 550 G, the components of $\vec{B_d}$ are 12.0~G along and 2.0 G transverse to the NV-center symmetry axis, respectively. Thus, at high microwave drive we can expect changes in the direction of $\vec{B_t}$ of at least 0.2$^\circ$. Reducing the dipolar field results in a total field that is more out-of-plane. This explains the contrast reversal in \ref{fig:Fig S3}(a). At negative values of $\theta_t$ the reduction of $\vec{B_d}$ results in an increase of the PL, and at positive values of $\theta_t$ the reduction of $\vec{B_d}$ results in a decrease of the PL, with the sign change of the response occurring near $\theta_t = 0^\circ$. The maximum slope of the PL versus applied field given in Figure \ref{fig:Fig S3}(c) suggests a 6\% change in PL per degree of field misalignment. The PL amplitudes in Figure \ref{fig:Fig S3}(a) suggest a static field angle suggest a magnetic field angle change of approximately 0.3$^\circ$, in agreement with our expected field angle changes given the dipolar field components.

\subsection{Micromagnetic Simulations}\label{sec:Appendix B3}

\begin{figure}[t]
\includegraphics[scale=1]{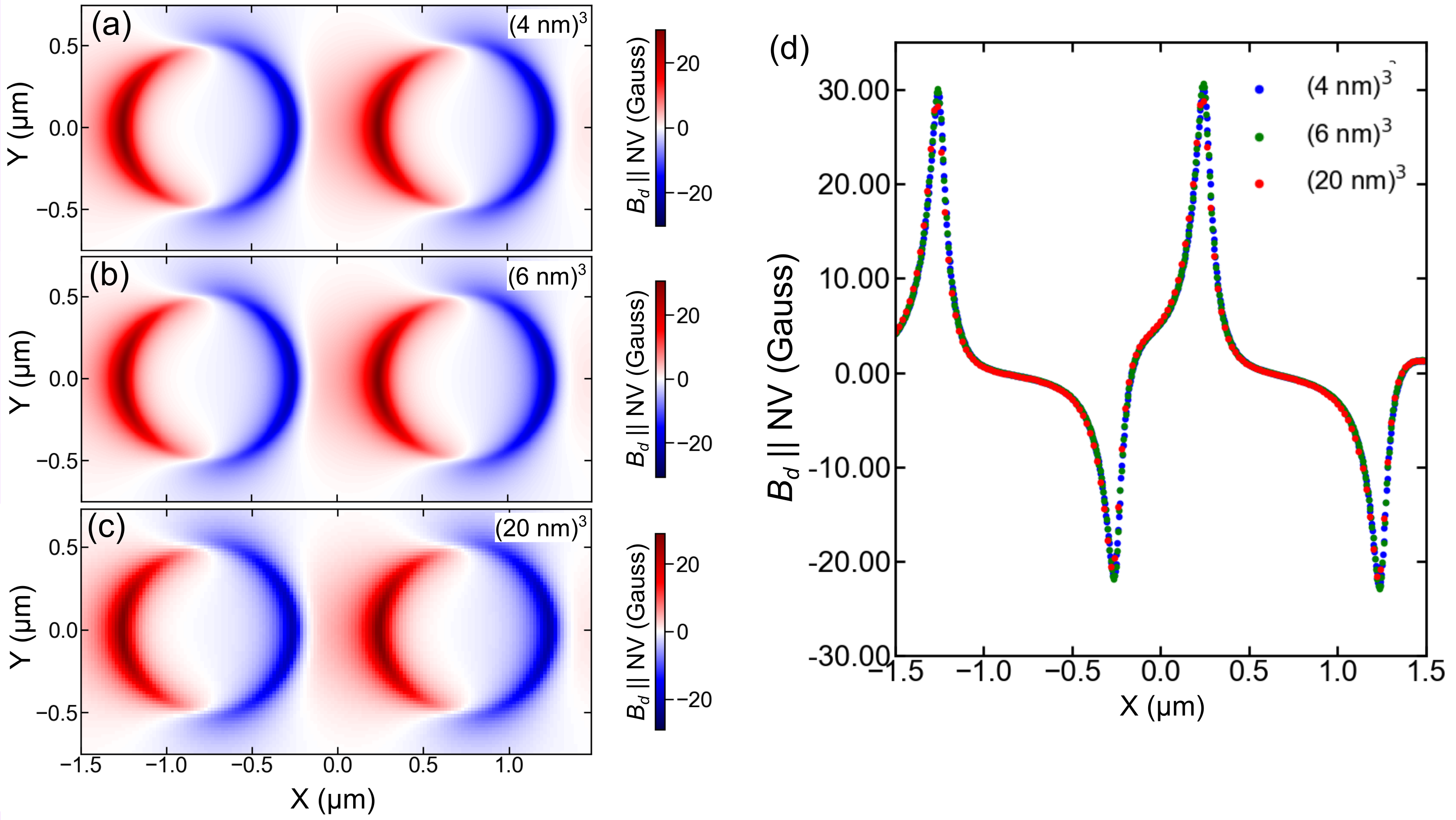}
\caption{\label{fig:Fig S4} Cell volume scaling of micromagnetic V[TCNE]$_{x \sim 2}$ disk pair simulations. Cross-section of $\vec{B_d}$ along the NV center symmetry axis $h = 50$ nm beneath a disk pair with  volumes of (a) (4 nm)$^3$ (b) (6 nm)$^3$ and (c) (20 nm)$^3$. Disk dimensions are $t = $ 100 nm, $d = 1 \unit{\um}$, and $s = 500$ nm. (d) Linecuts of the $\vec{B_d}$ along the NV center symmetry axis confirm that no simulation artifacts are present at cell volume (20 nm)$^3$, which is the smallest cell size that we are able to use for simulating $\vec{B_a}$ in our experimental geometry.}
\end{figure}

We use the mumax3 micromagnetic simulation package to simulate the dipolar fields of the disk pair~\cite{Vansteenkiste_2014}. In micromagnetics, the maximum allowable cell dimension is typically 75\% of the exchange length, which for V[TCNE]$_{x \sim 2}$ has been reported to be anywhere from 9.7 nm \cite{Franson_2019} to 21 nm \cite{Zhu_2016}. Our disk pair volume is sufficiently large that we are unable to simulate the entire sample using a strictly appropriate cell volume size. Thus, we perform a cell volume scaling test to determine how using a cell volume of (20 nm)$^3$, the smallest cell size for which we have enough GPU memory to store the entire simulation volume, influences the simulated dipolar fields. The results are shown in Figure \ref{fig:Fig S4} and confirm that no significant artifacts arise using a cell volumes up to (20 nm)$^3$. The component of the dipolar field at a height $h = 50$ nm beneath center point of the disk pair is about 4\% different than the value found when using a cell volume of (4 nm)$^3$.

\begin{figure}[t]
\includegraphics[scale=1]{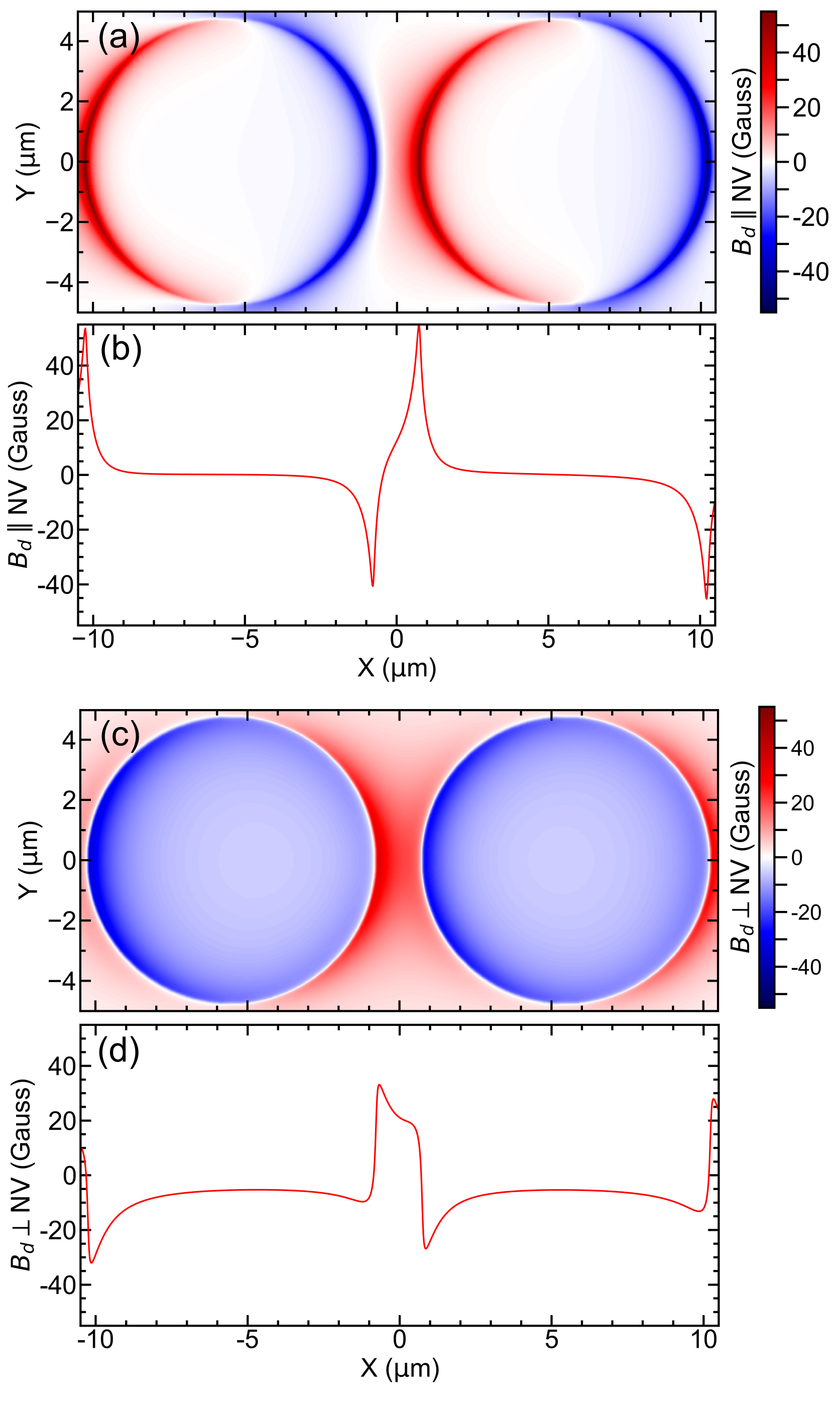}
\caption{\label{fig:Fig S5} Micromagnetic simulations of V[TCNE]$_{x \sim 2}$ disk pair for $\vert \vec{B_a} \vert = 641$ G and $\theta_a = $ 2.4$^\circ$. (a) Cross-section of the projection of $\vec{B_d}$ along the NV center symmetry axis $h = 50$ nm beneath the disks. (b) Linecut of panel (a) along the pair long axis. (c) Cross-section of the projection of $\vec{B_d}$ transverse to the NV center symmetry axis $h = 50$ nm beneath the disks. (b) Linecut of panel (c) along the pair long axis. Such simulations are used to determine the components of $\vec{B_d}$ and $\vec{B_a}$ for a known set of $\{ \vert \vec{B_t} \vert ,\theta_t \}$ for each curve in Figure \ref{fig:Fig 4}.}
\end{figure}

We include an example of the micromagnetic simulations for our experimental sample geometry used to find the values of $\vec{B_d}$ and $\vec{B_a}$ throughout our work. Such an example is shown in Figure \ref{fig:Fig S5}. The parameters are as follows: $\vert\vec{B_a}\vert = 641$~G, $\theta_a = 2.4^\circ$, $t = 700$~nm, $d = 9.5~\unit{\um}$, $s = 1.5~\unit{\um}$, the saturation magnetization $M_{\textrm{sat}} = 9549$~A/m \cite{Yu_2014}, and the exchange stiffness $A_{\textrm{ex}} = 2.2 \times 10^{-15}$~J/m \cite{Franson_2019}. The value of $M_{\textrm{sat}}$ at 200 K is taken from the DC magnetometry results performed on V[TCNE]$_{x \sim 2}$ thin films reported in reference \cite{Yu_2014}. The total simulation volume is $21~\unit{\um} \times 10~\unit{\um} \times 800 $ nm. We assume no uniaxial magnetic anisotropy. Since the applied field is at least an order of magnitude larger than the reported anisotropy fields in V[TCNE]$_{x \sim 2}$, the anisotropy has a negligible contribution to the stray dipolar field. In other words, the magnetic moments within the disk pair are saturated along the applied field direction.

\end{appendix}

\clearpage

\bibliography{bibliography}

\end{document}